\documentclass[a4paper,12pt]{article}

\newcommand{\sect}[1]{\setcounter{equation}{0}\section{#1}}

\begin{document}
\topmargin 0pt
\oddsidemargin 0mm

\renewcommand{\thefootnote}{\fnsymbol{footnote}}
\begin{titlepage}
\begin{flushright}
hep-th/0111093
\end{flushright}

\vspace{5mm}
\begin{center}
{\Large \bf Cardy-Verlinde Formula and Asymptotically de Sitter Spaces}
\vspace{12mm}

{\large
Rong-Gen Cai\footnote{email address: cairg@itp.ac.cn}}\\
\vspace{8mm}
{\em  Institute of Theoretical Physics, Chinese Academy of Sciences, \\
   P.O. Box 2735, Beijing 10008, China} 
\end{center}
\vspace{5mm}
\centerline{{\bf{Abstract}}}
\vspace{5mm}
In this paper we discuss the question of whether the entropy of cosmological horizon in 
some asymptotically de Sitter spaces can be described by  the Cardy-Verlinde formula, 
which is supposed to be an entropy formula of conformal field theory in any dimension.
For the Schwarzschild-de Sitter solution, although the gravitational mass is always
negative (in the sense of the prescription in hep-th/0110108 to calculate the
conserved charges of asymptotically de Sitter spaces), we find that indeed the entropy 
of cosmological horizon can be given by using naively the Cardy-Verlinde formula. The 
entropy of pure de Sitter spaces can also be expressed by the Cardy-Verlinde formula. 
For the topological de Sitter solutions, which have a cosmological horizon and a naked 
singularity, the Cardy-Verlinde formula also works well. Our result is in favour of the
dS/CFT correspondence. 

\end{titlepage}

\newpage
\renewcommand{\thefootnote}{\arabic{footnote}}
\setcounter{footnote}{0}
\setcounter{page}{2}

\sect{Introduction}

In a recent paper~\cite{Verl}, Verlinde argued that the Cardy 
formula~\cite{Cardy}, describing the entropy of a certain conformal field
theory (CFT) in $1+1$ dimensions, can be generalized to any dimension, leading
to the so-called Cardy-Verlinde formula.
 Consider a certain CFT
residing in an $(n+1)$-dimensional spacetime with the metric 
\begin{equation}
\label{1eq1}
ds^2 =-dt^2 +R^2d\Omega_n^2,
\end{equation}
where $d\Omega^2_n$ denotes the line element of a unit $n$-dimensional sphere and
$R$ is the radius of the sphere.
It was proposed that the entropy of the CFT in the spacetime (\ref{1eq1}) 
can be related to its total
energy $E$ and Casimir energy $E_c$ as
\begin{equation}
\label{1eq2}
S=\frac{2\pi R}{\sqrt{ab}}\sqrt{E_c(2E-E_c)}.
\end{equation}
Here $a$ and $b$ are two positive parameters. For strongly coupled CFTs with
the AdS duals, which implies that the CFTs are in the regime of supergravity
duals, $ab$ is fixed to be $n^2$ exactly. Thus one obtains the Cardy-Verlinde
formula~\cite{Verl}
\begin{equation}
\label{1eq3}
S=\frac{2\pi R}{n}\sqrt{E_c(2E-E_c)}.
\end{equation}
Indeed, this formula holds for various kinds of asymptotically AdS spacetimes whose
boundary metric is of the form (\ref{1eq1}):
AdS Schwarzschild black holes~\cite{Verl}; 
charged AdS black holes~\cite{Cai1} and  Taub-Bolt AdS spacetimes~\cite{Birm},
whose thermodynamics corresponds to that of different CFTs~\cite{Witten2}. The 
Cardy-Verlinde formula (\ref{1eq3}) holds even for the AdS Kerr black 
holes~\cite{Klem1}, whose boundary is a rotating Einstein universe.\footnote{For
more references discussing the Cardy-Verlinde formula see \cite{Kuta}-\cite{Brev}.}

More recently, it has been proposed that defined in a manner analogous to the 
AdS/CFT correspondence,  quantum gravity in a de Sitter (dS) space is dual to a certain
 Euclidean  CFT living on a spacelike boundary of the 
dS space~\cite{Strom} (see also earlier works \cite{Hull}-\cite{Witten1}). Following
the proposal, some investigations on the dS space have been carried out 
recently~\cite{Mazu}-\cite{Ogus}. 

It is well known that in de Sitter space, there is a cosmological horizon, which
has the similar thermodynamic properties like the black hole horizon~\cite{Gibbons}.
According to the dS/CFT correspondence, it might be expected that as the case of AdS
black holes~\cite{Witten2}, the thermodynamics of cosmological horizon in asymptotically
dS spaces can be identified with that of a certain Euclidean CFT residing on a 
spacelike boundary of the asymptotically dS spaces. Thus it is of great interest to see
whether the entropy of the cosmological horizon can be described by the Cardy-Verlinde
formula (\ref{1eq3}). This is the purpose of the present paper.\footnote{see \cite{Dani}
for another attempt in this direction. The dynamics of a brance in  a 
Schwarzschild-dS spacetime has been discussed in \cite{Ogus}.}

To this end, we have to first face a difficulty to calculate some conserved charges 
including the mass (total energy) of gravitational field of the asymptotically dS 
spaces. In the spirit of the dS/CFT correspondence, these conserved charges of 
gravitational field can be identified with those of corresponding Euclidean CFT. The 
difficulty arises due to the absence of  the spatial infinity and the globally timelike
Killing vector in an asymptotically dS space. In this paper we will follow the 
prescription recently proposed in \cite{BBM} to calculate the mass of gravitational field 
of asymptotically dS spaces (and then the energy of corresponding CFTs). It is found 
that indeed the entropy of cosmological horizon in some asymptotically dS spaces can 
be described in terms of the Cardy-Verlinde formula.

The organization of the paper is as follows. In the next section we first discuss
the case of Schwarzschild-dS solutions. In Sec.~3 we will consider the case of topological
dS solutions presented in \cite{CMZ}. We conclude with a discussion in Sec.~4.

\sect{Schwarzschild-dS Solutions}

Consider an $n+2$-dimensional Schwarzschild-dS 
spacetime\footnote{In three dimensions, it has been already shown that the entropy
of Schwarzschild (Kerr)-de Sitter solution can be expressed by Cardy formula, for
example see \cite{BBM} and references therein. So in this paper we consider the case
of spacetime dimension $n+2 \ge 4$.}
\begin{equation}
\label{2eq1}
ds^2 = -f(r) dt^2 +f(r)^{-1}dr^2 +r^2 d\Omega_n^2,
\end{equation}
where 
\begin{equation}
f(r) = 1 -\frac{\omega_n m}{r^{n-1}} -\frac{r^2}{l^2}, \ \ \ 
\omega_n=\frac{16\pi G}{nVol(S^n)}.
\end{equation}
Here $m$ is an integration constant, $Vol(S^n)$ denotes the volume of a unit $n$-sphere 
$d\Omega_n^2$, and $G$ is the gravitational constant. 
The Schwarzschild-dS spacetime (\ref{2eq1}) is a solution of Einstein equations with 
a cosmological constant $\Lambda =n(n+1)/2l^2$ in $n+2$ dimensions. When the 
Schwarzschild parameter $m$ vanishes, the solution (\ref{2eq1}) reduces to the
dS solution, which has a cosmological horizon at $r=l$. The horizon has the associated 
Hawking temperature $T_{\rm HK}=1/2\pi l$ and entropy $S=l^nVol(S^n)/4G$~\cite{Gibbons}. When 
$m$ increases with $m>0$, a black hole horizon occurs and increases in size with $m$, 
while the cosmological horizon shrinks. Finally the black hole horizon and cosmological
horizon coincide with each other when 
\begin{equation}
\label{2eq3}
m_N = \frac{2}{\omega_n (n+1)}\left(\frac{n-1}{n+1}l^2\right)^{(n-1)/2}.
\end{equation}
This is the Nariai black hole, the maximal black hole in dS space.  

When $m\ne 0$, the cosmological horizon $r_c$ of the solution (\ref{2eq1}) is 
determined by the maximal root of the equation $f(r)=0$. The associated Hawking temperature
and entropy to the cosmological horizon are
\begin{eqnarray}
\label{2eq4}
&& T_{\rm HK} =\frac{1}{4\pi r_c}\left ((n+1)\frac{r_c^2}{l^2}-(n-1)\right),
   \nonumber \\
&& S =\frac{r_c^n Vol(S^n)}{4G}.
\end{eqnarray}
In Ref.~\cite{BBM}, using surface counterterm method the authors calculated the mass 
of gravitational field  of the spacetime (\ref{2eq1}) in four and five 
dimensions~\footnote{For the case in  higher dimensions, more counterterms than those
given in \cite{BBM} are needed.}. It was found that
\begin{equation}
\label{2eq5}
M_4 =-m, \ \ \  M_5=\frac{3\pi l^2}{32G} -m.
\end{equation}
Thus a pure dS$_4$ has a vanishing mass, while the Nariai black hole has the mass
$M_4 =-l/3\sqrt{3}G$ in four dimensions. 
The pure dS$_5$ has the mass $3\pi l^2/32G$ and the Nariai
black hole in five dimensions has a vanishing mass. This is a surprising result: The
mass of black hole solution is always less than that of pure dS space in corresponding
dimensions. However, it was argued that this result is consistent with the Bousso's
observation on the asymptotically dS space and the dS/CFT correspondence~\cite{BBM}.
The Bousso's observation is that the entropy of dS space is an upper bound for the entropy
of any asymptotically dS spaces~\cite{Bous}. Furthermore, if the dS/CFT correspondence
exists, the mass of asymptotically dS spaces should be translated into the energy of 
a dual Euclidean CFT. Generically it is expected that such a field theory has entropy 
increasing with energy. The result (\ref{2eq5}) has precisely this 
property~\cite{BBM}.  

The nonvanishing mass of pure dS$_5$ space is reminiscent of the nonvanishing
mass of pure AdS$_5$ in the global coordinates~\cite{Bala2}. The latter is shown to be 
just
the Casimir energy of ${\cal N}=4$ SYM theory on a $3$-sphere. Such a Casimir energy
is expected to exist for pure AdS space in odd dimensions~\cite{Bala2}. In the closest
analogy of the case in AdS spaces, the nonvanishing mass of pure dS spaces is also
expected to exist in odd dimensions in some coordinates ({\it cf.} \cite{CMZ}). These
nonvanishing masses can be viewed as the Casimir energies of CFTs dual to the pure
dS spaces. 

Note that in checking the Cardy-Verlinde formula for the asymptotically AdS 
spaces, the Casimir energy of pure AdS spaces did not be 
included~\cite{Verl,Cai1,Birm,Klem1}\footnote{If one wants to include the Casimir
energy of the pure AdS space, this part must be subtracted from the total energy in 
the Cardy-Verlinde formula, in a way as we 
did in the case of charged AdS black holes~\cite{Cai1}.}.
 In a manner analogous to this, we will not
also consider the Casimir energy corresponding to the pure dS spaces in the present
context. As a result, we obtain the gravitational mass of the 
Schwarzschild-dS solution (\ref{2eq1}) from (\ref{2eq5})~\footnote{For other higher 
dimensions than $4$ and $5$, we expect
that the following result holds also.}
\begin{equation}
\label{2eq6}
E=-m =\frac{r_c^{n-1}}{\omega_n}\left(\frac{r_c^2}{l^2}-1\right),
\end{equation}
where $r_c$ denotes the cosmological horizon. According to the dS/CFT correspondence,
the gravitational mass (\ref{2eq6}) is just the energy of the dual CFT. Thus,
 in this case the energy of CFT is negative, a unpleasant result\footnote{We have more
discussions on this below.}. Despite of the non-positive definiteness of the energy, 
 we first follow naively the steps as the case of AdS black holes to check whether
the entropy (\ref{2eq4}) of the cosmological horizon can be expressed or not by the
Cardy-Verlinde formula (\ref{1eq3}).

Following Ref.~\cite{Verl}, the Casimir energy $E_c$ can be calculated using the
formula 
\begin{equation}
\label{2eq7}
E_c=n(E +pV -T_{\rm HK}S),
\end{equation}
where $p$ is the pressure and $V$ is the volume. Since we are considering a CFT, 
so we have $p=E/nV$. Corresponding to the
Schwarzschild-dS solution (\ref{2eq1}), the dual Euclidean CFT resides on the space
\begin{equation}
\label{2eq8}
ds^2 =dt^2 +l^2 d\Omega_n^2. 
\end{equation}
Note that here $t$ denotes a spacelike coordinate, rather than a timelike coordinate.
Substituting (\ref{2eq4}) and (\ref{2eq6}) into (\ref{2eq7}), we obtain
\begin{equation}
\label{2eq9}
E_c=-\frac{2nr_c^{n-1}Vol(S^n)}{16\pi G}.
\end{equation}
This Casimir energy is negative, which is reminiscent of the case of the hyperbolic
AdS black holes, there the Casimir energy is also negative~\cite{Cai1}. The negative
Casimir energy indicates that the dual CFT is not unitary. This agrees with
Ref.~\cite{Strom}, in which it was argued that the dual Euclidean CFT to the 
dS space is not unitary.

Given the Casimir energy (\ref{2eq9}), it is easy to check that the entropy $S$
(\ref{2eq4}) can be expressed as follows
\begin{equation}
\label{2eq10}
S =\frac{2\pi l}{n}\sqrt{|E_c|(2E-E_c)}. 
\end{equation}
This form is completely the same as the Cardy-Verlinde formula in the case of
hyperbolic AdS black holes~\cite{Cai1}. Note that this formula (\ref{2eq10}) also
holds for pure dS spaces. In that case, $E=0$ and 
\begin{equation}
\label{2eq11}
E_c=-\frac{2nl^{n-1}Vol(S^n)}{16\pi G},
\end{equation}
the expression (\ref{2eq10}) then gives the entropy of pure dS spaces.
The result (\ref{2eq10}) looks fine. But there are two points to be understood. The 
first is that in this formula the energy of corresponding CFT is 
negative\footnote{If replacing $E=-m$ by $E=m$ in (\ref{2eq6}), one then cannot arrive
at the Cardy-Verlinde formula (\ref{2eq10}).}. The other is that the 
formula (\ref{2eq10}) gives us the entropy of cosmological horizon only, and does not
apply to the black hole horizon. As we stated above, when $m\le m_N$, except for the
cosmological horizon, a black hole horizon $r_+$ occurs, which has the Hawking
temperature $\tilde{T}_{\rm HK}$ and entropy $\tilde{S}$
\begin{eqnarray}
&& \tilde{T}_{\rm HK}=\frac{1}{4\pi r_+}\left((n-1)-(n+1)\frac{r_+^2}{l^2}\right),
   \nonumber \\
&& \tilde{S}=\frac{r_+^nVol(S^n)}{4G}.
\end{eqnarray}
For the Schwarzschild-dS solution, the total entropy should be the sum of black
hole horizon entropy $\tilde{S}$ and cosmological horizon entropy $S$. But we cannot
obtain a similar formula like (\ref{2eq10}) for the entropy $\tilde{S}$ 
of black hole horizon. Therefore further investigation is needed for the Schwarzschild-dS
spacetime. In order to avoid the two points, in the next section we consider the topological
dS solutions presented in~\cite{CMZ}.

\sect{Topological dS Solutions}

In Ref.~\cite{BBM}, an interesting conjecture was put forward, which states that
{\it any asymptotically de Sitter space whose mass exceeds that of de Sitter contains
a cosmological singularity}. To check this conjecture, in Ref.~\cite{CMZ} 
 Myung, Zhang and the present author present a solution named topological de Sitter
solution. The solution can be described by the metric
\begin{equation}
\label{3eq1}
ds^2 = -f(r)dt^2 +f(r)^{-1}dr^2 +r^2 \gamma_{ij}dx^idx^j,
\end{equation}
where
\begin{equation}
\label{3eq2}
f(r)=k+\frac{\omega_nm}{r^{n-1}} -\frac{r^2}{l^2},\ \ \omega_n=\frac{16\pi G}
  {n Vol(\sigma)},
\end{equation}
$m$ is an integration constant which is supposed to be positive, and 
$\gamma_{ij}dx^idx^j$ denotes the line element of an $n$-dimensional 
 hypersurface $\sigma$ with constant curvature $n(n-1)k$ and volume $Vol(\sigma)$.
Without loss of generality, the constant $k$ can be set to $1$, $0$ or $-1$.

The solution (\ref{3eq1}) is asymptotically de Sitter. But because of $m\ge 0$, 
black hole horizon is not present in this solution, instead a naked singularity
at $r=0$ occurs when $m \ne 0$.  Although so, the solution has a cosmological 
horizon $r_c$, which is the root of the equation, $f(r)=0$. And the associated 
Hawking temperature $T_{\rm HK}$ and entropy $S$ to the cosmological horizon are
\begin{eqnarray}
\label{3eq3}
&& T_{\rm HK}= \frac{1}{4\pi r_c}\left((n+1)\frac{r_c^2}{l^2}-(n-1)k\right),
   \nonumber \\
&& S =\frac{r_c^n Vol(\sigma)}{4G}.
\end{eqnarray}

When $k=1$, the solution (\ref{3eq1}) has the same form as that of the Schwarzschild-dS
solution if $m$ is replaced by $-m$ in (\ref{2eq1}). Due to this, following the
prescription in Ref.~\cite{BBM}, it is easy to show
that the gravitational masses of the solutions (\ref{3eq1}) in four and five dimensions
are
\begin{equation}
\label{3eq4}
M_4 =m, \ \ \  M_5 = \frac{3\pi l^2}{32 G}+m,
\end{equation}
respectively. When $k=0$, the gravitational mass was found to be~\cite{CMZ}
\begin{equation}
\label{3eq5}
M=m, 
\end{equation}
while when $k=-1$, the gravitational masses in four and five dimensions are~\cite{CMZ}
\begin{equation}
\label{3eq6}
M_4=m, \ \ \  M_5 =\frac{3l^2 Vol(\sigma)}{64\pi G} +m.
\end{equation}
For these cases, the gravitational masses are always larger than those
of de Sitter spaces ($m=0$) in corresponding dimensions. As a result, we verify
the conjecture in Ref.~\cite{BBM} within these examples. For more details 
see \cite{CMZ}. 

Corresponding to the solution (\ref{3eq1}), the dual Euclidean CFT resides on the
space 
\begin{equation}
\label{3eq7}
ds^2 =dt^2 +l^2 \gamma_{ij}dx^idx^j. 
\end{equation}
Once again, here the coordinate $t$ is a spacelike one. As the case of asymptotically
AdS spaces, we neglect the nonvanishing mass of pure dS spaces, as those in (\ref{3eq4})
and (\ref{3eq6}), or subtract the nonvanishing energy from the total energy of 
corresponding CFTs, the energy of the Euclidean CFTs dual to the solution (\ref{3eq1})
is then 
\begin{equation}
\label{3eq8}
E=m =\frac{r_c^{n-1}}{\omega_n}\left( \frac{r_c^2}{l^2}-k\right).
\end{equation}
Note that here the energy of CFTs is always positive, different from the case 
(\ref{2eq6}) of the Schwarzschild-dS solution. Substituting (\ref{3eq3}) and 
(\ref{3eq8}) into (\ref{2eq7}), we get the Casimir energy $E_c$
\begin{equation}
\label{3eq9}
E_c=-\frac{2nk r_c^{n-1}Vol(\sigma)}{16\pi G}.
\end{equation}
When $k=0$, the Casimir energy vanishes, as the case of asymptotically AdS spaces.
This is expected since the thermodynamic quantities of CFTs in a Ricci flat space are
conformal invariant, there is no finite volume effect.  When $k=\pm 1$, we see
from (\ref{3eq9}) that the sign of the energy is just contrast to the case of 
asymptotically AdS space~\cite{Cai1}, there it was found that for a hyperbolic space
with $k=-1$, the Casimir energy is negative~\cite{Cai1}, while it is positive
for the $k=1$ case~\cite{Verl}. 

With the Casimir energy (\ref{3eq9}), one can easily see  that the entropy (\ref{3eq3})
of the cosmological horizon can be expressed in a form of the Cardy-Verlinde formula
as 
\begin{equation}
\label{3eq10}
S =\frac{2\pi l}{n}\sqrt{|E_c|(2E-E_c)},
\end{equation}
when $k=\pm 1$. To accommodate the case of $k=0$, we can rewrite (\ref{3eq10}) as 
\begin{equation}
\label{3eq11}
S=\frac{2\pi l}{n}\sqrt{|E_c/k|(2E-E_c)}.
\end{equation}
As a result, we show that indeed the entropy of the cosmological horizon in the 
topological dS solution (\ref{3eq1}) can be described by the Cardy-Verlinde 
formula (\ref{1eq3}). This also further provides evidence in favour of the dS/CFT
correspondence.


\sect{Conclusions}

In the dS/CFT correspondence, we have investigated the question of whether the entropy
of cosmological horizon in asymptotically dS spaces can be described by the
Cardy-Verlinde formula, which was established in the AdS/CFT 
correspondence~\cite{Verl}.  For the Schwarzschild-dS solution, although the gravitational
mass, calculated in the prescription of Ref.~\cite{BBM}, of the solution (and then the 
energy of the dual Euclidean CFT) is always negative, we have found that the 
entropy of the cosmological horizon (and then the entropy of the dual CFT) indeed can 
be expressed in terms of a form [see (\ref{2eq10})] of the Cardy-Verlinde formula. However,
we cannot find a similar formula for the entropy of black hole horizon in the 
Schwarzschild-de Sitter spacetime. Therefore further study is needed for the 
Schwarzschild-dS solution. When the Schwarzschild parameter $m$ vanishes, our result 
(\ref{2eq10}) precisely reproduces the entropy of pure dS spaces.

In the topological dS solutions~\cite{CMZ}, which have a cosmological horizon and a 
naked singularity at $r=0$. The gravitational mass is always positive. The entropy 
associated to the cosmological horizon was found to obey the Cardy-Verlinde formula
[see (\ref{3eq11})]. This result is in favour of the dS/CFT correspondence. One point
which should be further investigated is whether the Euclidean CFT dual to the topological
dS solution can describe  the naked singularity in this asymptotically dS 
spacetime. In other words, whether is there a well-defined Euclidean CFT dual to an 
asymmetrically dS space which contains a naked singularity?

It would be of interest to extend the discussions made in this paper to the case of 
asymptotically dS spaces with rotation and/or charges.

\section*{Acknowledgments}
The author thanks Y.S. Myung for useful discussions. 
This work was supported in part by a grant from Chinese Academy of Sciences.

\newpage

\end{document}